\documentclass[11pt, reqno]{article}%{amsart}
\usepackage{latexsym, amsfonts, amsthm, amscd}
\usepackage{amssymb, amsmath}
\usepackage{hyperref}

	\theoremstyle{plain}
	\newtheorem{prop}{Proposition}[section]
        \newtheorem{theorem}[prop]{Theorem}

	\theoremstyle{definition}
	
	\newtheorem{definition}{Definition}[section]

	\theoremstyle{remark}

	\newcommand{\Real}{\mathbb{R}}

	\newcommand{\Ocal}{\mathcal{O}}

	\newcommand{\beq}{\begin{eqnarray}}
	\newcommand{\eeq}{\end{eqnarray}}
	\newcommand{\beqs}{\begin{eqnarray*}}
	\newcommand{\eeqs}{\end{eqnarray*}}

	\newcommand{\bthm}{\begin{theorem}}
	\newcommand{\ethm}{\end{theorem}}

\newcommand{\ldiag}[1]%
       {\makebox[0cm]{${\scriptstyle#1}\downarrow\phantom{\scriptstyle#1}$}}
\newcommand{\ldiagup}[1]%
       {\makebox[0cm]{${\scriptstyle#1}\uparrow\phantom{\scriptstyle#1}$}}
\newcommand{\rdiag}[1]%
       {\makebox[0cm]{$\phantom{\scriptstyle#1}\downarrow{\scriptstyle#1}$}}
\newcommand{\sediagr}[1]%
       {\makebox[0cm]{$\phantom{\scriptstyle#1}\searrow{\scriptstyle#1}$}}
\newcommand{\nediagr}[1]%
       {\makebox[0cm]{$\phantom{\scriptstyle#1}\nearrow{\scriptstyle#1}$}}
\newcommand{\rdiagup}[1]%
       {\makebox[0cm]{$\phantom{\scriptstyle#1}\uparrow{\scriptstyle#1}$}}
\newcommand{\swdiag}[1]%
       {\makebox[0cm]{$\phantom{\scriptstyle#1}\swarrow{\scriptstyle#1}$}}
\newcommand{\sediag}[1]%
       {\makebox[0cm]{${\scriptstyle#1}\searrow\phantom{\scriptstyle#1}$}}
\newcommand{\nediag}[1]%
       {\makebox[0cm]{${\scriptstyle#1}\nearrow\phantom{\scriptstyle#1}$}}
\newcommand{\code}[1]{{\tt #1}}

\begin{document}
\title{An efficient algorithm to find a set of nearest elements in a mesh}

\author{Gleb Novitchkov
\\
%The 21st Century COE Program, \\
%Center for Integrative Mathematical Sciences,\\
%Keio University,\\
%3-14-1 Hiyoshi, Kohoku-ku, Yokohama 223-8522, Japan\\
\\
{\sf email: gleb.novitchkov@gmail.com }}                                                         
\maketitle
\tableofcontents
%\acknowledgments 
%\begin{abstract}
%\end{abstract}

\section{Introduction}
Here we present an algorithm that find a list of elements neighboring some given element in a linear time.
More precisely, if there are $N_{\rm elem}$ elements in the mesh, the runtime of the algorithm is 
$\Ocal(N_{\rm elem})$.

\section{Definition}
\begin{definition}
By \emph{element} we mean a 3-simplex $\Delta^3$ imbedded in $\Real^3$.
\end{definition}
Essentially an element is a tetrahedron in $\Real^3$.

Goal:  given an element of the mesh, we want to find a set of elements that are near this element.
\begin{definition}
Given an element $E$, we call an element $F$ a {\bf vertex-near} element of $E$ 
if $E$ and $F$ share common vertex; we call an element $F$ a {\bf edge-near} element of $E$ 
if $E$ and $F$ share common edge; we call an element $F$ a {\bf face-near} element of $E$ 
if $E$ and $F$ share common face.  An element $F$ is {\bf near }element of $E$ if $F$ is either face-near, edge-near,
of vertex-near element of $E$.
\end{definition}

\subsection{Mesh and the representation of the elements}
A mesh is specified by the cloud of $n$ points, or \emph{nodes}, $\{p_0, p_1,\dots, p_{n-1}\}$, where $p_i = (x_i, y_i, z_i)$.
An element $E_i$ is specified by the four nodes, $E_i = \{p_{i_0}, p_{i_1},p_{i_2},p_{i_3}\}$.

\section{The algorithm}
The idea of the algorithm is to do the histogramming of the nodes.
\begin{description}
\item[Step 1.]  (initialization) For $N_{\rm node}$ nodes allocate array $L$ of lists, the length of the array is $N_{\rm node}$.  
\item[Step 2.]  (histogramming) For each element $i_{\rm elem}$, $0\leq i_{\rm elem}\leq N_{\rm elem}-1$, do:
\begin{description}
	\item[Step 2.1]  For each node $p_{j_i}$, $0\leq i\leq 3$ of element $i_{\rm elem}$ add index 
	$i_{\rm elem}$ to the lists $L[{j_0}]$, $L[{j_1}]$, $L[{j_2}]$, $L[{j_3}]$.  In C++ notation, 
	it should be: 
		\code{L[$j_0$].push\_back($i_{\rm elem}$)}, 
		\code{L[$j_1$].push\_back($i_{\rm elem}$)}, 
		\code{L[$j_2$].push\_back($i_{\rm elem}$)}, 
		\code{L[$j_3$].push\_back($i_{\rm elem}$)}.
\end{description}
\item[Step 3.]  (Finding neighboring elements) Given element $E_i = (j_0, j_1, j_2, j_3)$, list of
elements neighboring $E_i$ is given by the union of the lists  $L[{j_0}]$, $L[{j_1}]$, $L[{j_2}]$, and $L[{j_3}]$.

\item[Step 4.] Deallocate array $L$ of lists.

\end{description}

\section{Parallelization}
Algorithm admits easy parallelization by OpenMP, one only has to take care to use critical section for Step 2.1:
adding element index to the lists should be done inside the critical section.

\section{Conclusion}
Missing details will be provided later.

\end{document}